\documentclass[reprint,aps,prb,superscriptaddress,notitlepage]{revtex4-1}

\usepackage{graphics,epsfig,amsfonts,amssymb,amsmath,color}
\usepackage[normalem]{ulem}
\usepackage{bbm}
\usepackage[mathcal]{euscript}
\usepackage{color}
\usepackage{float}
\usepackage{braket}

\begin{document}

\title{Signatures of atomic-scale structure in the energy dispersion and coherence of a Si quantum-dot qubit}
\author{J. C. Abadillo-Uriel}
\affiliation{Materials Science Factory, Instituto de Ciencia de Materiales de Madrid (ICMM), Consejo Superior de Investigaciones Cient\'ificas (CSIC), Sor Juana Ines de la Cruz 3, 28049 Madrid, Spain}
\affiliation{Department of Physics, University of Wisconsin-Madison, Madison, WI 53706, United States}
\author{Brandur Thorgrimsson}
\affiliation{Department of Physics, University of Wisconsin-Madison, Madison, WI 53706, United States}
\author{Dohun Kim}
\affiliation{Department of Physics and Astronomy, and Institute of Applied Physics, Seoul National University, Seoul 08826, South Korea}
\author{L. W. Smith}
\thanks{Current address: Cavendish Laboratory, Department of Physics, University of Cambridge, J. J. Thomson Avenue, Cambridge CB3 0HE, United Kingdom.}
\affiliation{Department of Physics, University of Wisconsin-Madison, Madison, WI 53706, United States}
\author{C. B. Simmons}
\affiliation{Department of Physics, University of Wisconsin-Madison, Madison, WI 53706, United States}
\author{Daniel R. Ward}
\affiliation{Department of Physics, University of Wisconsin-Madison, Madison, WI 53706, United States}
\author{Ryan H. Foote}
\author{J. Corrigan}
\affiliation{Department of Physics, University of Wisconsin-Madison, Madison, WI 53706, United States}
\author{D. E. Savage}
\author{M. G. Lagally}
\affiliation{Department of Materials Science and Engineering, University of Wisconsin-Madison, Madison, WI 53706, United States}
\author{M. J. Calder\'{o}n}
\affiliation{Materials Science Factory, Instituto de Ciencia de Materiales de Madrid (ICMM), Consejo Superior de Investigaciones Cient\'ificas (CSIC), Sor Juana Ines de la Cruz 3, 28049 Madrid, Spain}
\author{S. N. Coppersmith}
\author{M. A. Eriksson}
\author{Mark Friesen}
\affiliation{Department of Physics, University of Wisconsin-Madison, Madison, WI 53706, United States}

\begin{abstract}
\noindent\textbf{ABSTRACT:} We report anomalous behavior in the energy dispersion of a three-electron double-quantum-dot hybrid qubit and argue that it is caused by atomic-scale disorder at the quantum-well interface. By employing tight-binding simulations, we identify potential disorder profiles that induce behavior consistent with the experiments. The results indicate that disorder can give rise to ``sweet spots" where the decoherence caused by charge noise is suppressed, even in a parameter regime where true sweet spots are unexpected. Conversely, ``hot spots" where the decoherence is enhanced can also occur. 
Our results suggest that, under appropriate conditions, interfacial atomic structure can be used as a tool to enhance the fidelity of Si double-dot qubits.

\vspace{0.2cm}\noindent\textbf{KEYWORDS:} \emph{Quantum dot, quantum well, qubit, silicon, valley splitting, tunneling}
\end{abstract}

\maketitle

Group IV materials are promising hosts for spin qubits \cite{KaneNature1998,Vrijen2000,ZwanenburgRMP} due to the predominance of nuclear spin-0 isotopes,\cite{ItohMRS2014} and the consequent abatement of magnetic noise.
Electrical (``charge") noise remains a problem, however, and it is ubiquitous across materials platforms.\cite{HuPRL2006,TaylorPRB2007} Charge noise has been shown to affect quantum-double-dot qubits, principally through the detuning control parameter,\cite{DialPRL2013} resulting in dephasing that depends on the energy dispersion as a function of detuning \citep{PeterssonPRL2010}. For Si dots, this dispersion is strongly affected by the physics of the conduction band minima, or ``valleys."\cite{FriesenAPL2006,CulcerPRB2009} Notably, atomic-scale disorder at the quantum-well interface affects the valley-orbit coupling and the tunnel coupling between dots,\cite{FriesenAPL2006,NestoklonPRB2006,KharcheAPL2007,FriesenPRB2010, CulcerPRB2010, Wu2012, GamblePRB2013,RohlingPRL2014,BorossPRB2016,GambleAPL2016} and thus the qubit frequency.

Here we show that random, atomic-scale disorder at the quantum well interface, combined with the ability to electrostatically manipulate the dot positions, enables us to exploit ``sweet spots" in the energy dispersion, where the effects of charge noise are strongly suppressed.\cite{Vion886, TaylorPRL2013, MedfordPRL2013, KimNature2015, JiajiaPRB2015, ReedPRL2016, MartinsPRL2016, ShimPRB2016} Sweet spots occur when the derivative of the qubit frequency with respect to the detuning parameter vanishes, $\partial f_Q/\partial \varepsilon =0$, since in this case, small $\varepsilon$ fluctuations do not cause variations of $f_Q$.  We report experimental evidence for a sweet spot occurring in an unexpected regime of control space, as well as the converse effect where decoherence is strongly enhanced by a ``hot spot."\cite{YangNC2013}  We also provide potential explanations for these phenomena in the form of specific disorder profiles that generate similar energy dispersions in two-dimensional (2D) tight-binding simulations of a double-quantum dot in a SiGe/Si/SiGe quantum well.

We focus on a specific qubit implementation, the quantum-dot hybrid qubit,\cite{ShiPRL2012, KohPNAS2013, KimNature2014, Ferraro2014, DeMichielisJPA2015, CaoPRL2016, WongPRB2016, ChenPRB2017} which behaves as a charge qubit when the detuning is close to zero, and has a spin-like character for large detuning values, $\varepsilon\gg 0$. 
The double dot device used in this work was grown on a step-graded SiGe virtual substrate that was miscut 2$^\circ$ towards (010),\cite{Mooney:1996} with the gate structure shown in Figure~1a.
Details about the device and its operation are presented in refs~\onlinecite{KimNature2014,Kim:2015npj,BrandurNPJ2017}. 

\begin{figure}[t]
\includegraphics[width=2.9 in]{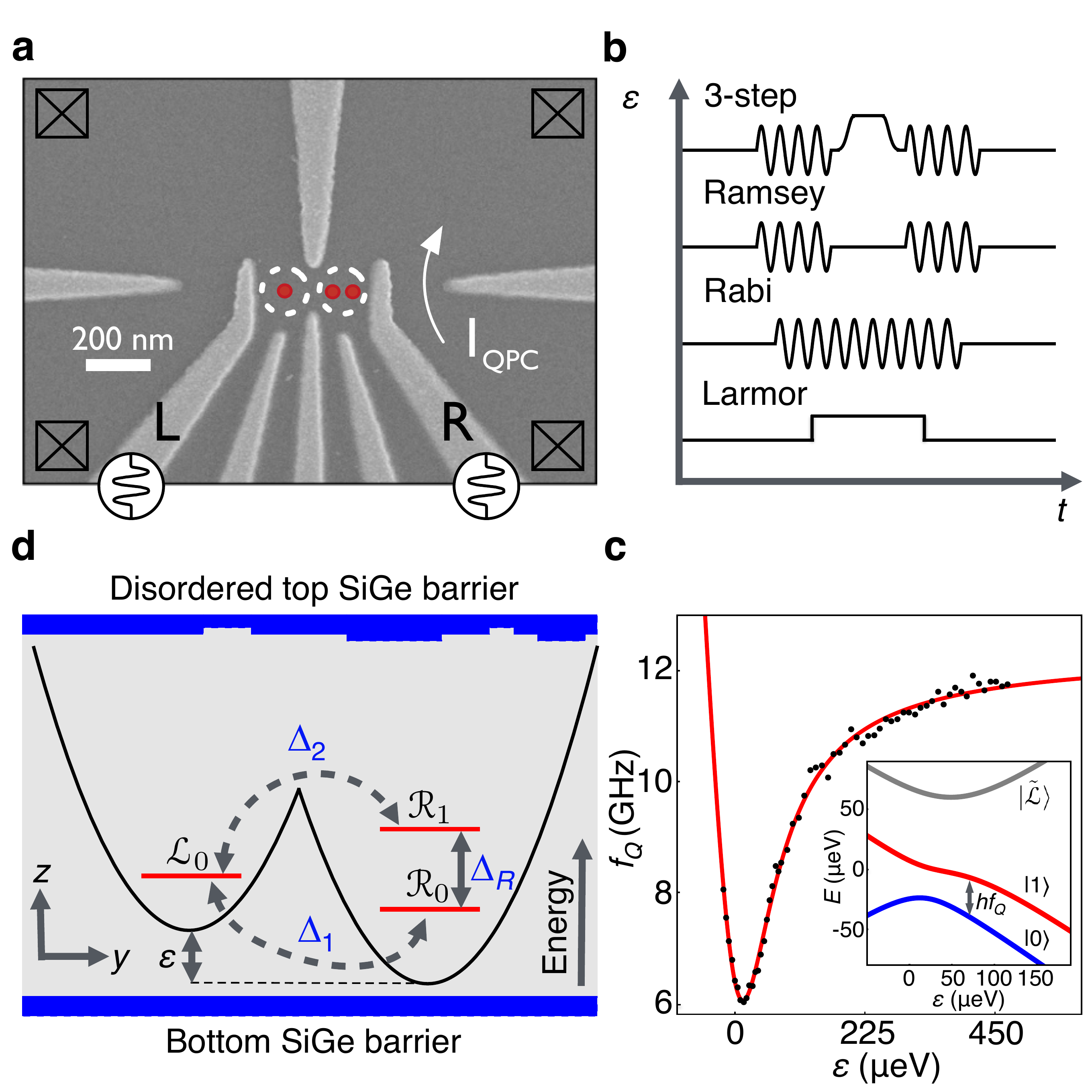}
\caption{
Experimental and theoretical set-up, and resulting energy dispersions.
(a) A scanning electron microscope image of a device nominally identical to the one used in the experiment. 
The gate voltages are tuned to form two quantum dots, located approximately within the dashed circles, where red dots represent electrons in a (1,2) charge configuration. 
(b) Schematics of the four pulse sequences employed in the experiments. 
The 3-step sequence is used to obtain the qubit frequency data, $f_Q$, plotted in (c). 
The Ramsey pulse sequence is used to obtain the qubit frequencies and Ramsey decay rates plotted in Figure~2. 
The Rabi and Larmor sequences are used to obtain Rabi fringes and $f_Q$ in Figure~3.
(c) The experimentally measured $f_Q$ of a quantum-dot hybrid qubit as a function of detuning, $\varepsilon$ (black dots). 
The solid red line shows the results of a least-squares fit of the data to the Hamiltonian, eq~1, assuming $\varepsilon$-independent model parameters.
Inset: the three energy eigenstates obtained by diagonalizing eq~1.
(d) A schematic cartoon illustrating the theoretical model for both the quantum-dot hybrid qubit and the single-electron charge qubit, with  the low-energy basis states, $\ket{\mathcal{L}_0}$, $\ket{\mathcal{R}_0}$, and $\ket{\mathcal{R}_1}$, as appropriate for the hybrid qubit. 
In our 2D tight-binding simulations, atomic-scale step disorder is introduced into the top interface as shown here and described in Methods.
The lateral confinement potential is taken to be biquadratic, and the two dots are offset by energy $\varepsilon$. 
The interdot tunnel couplings are labelled $\Delta_1$ and $\Delta_2$, and we refer to $\Delta_R$ as the ``valley splitting," although $\ket{\mathcal{R}_1}$ may involve a valley-orbit excitation.}
\label{fig1}
\end{figure} 

Here we employ four different pulse sequences to determine the qubit energy dispersion, as illustrated in Figure~1b and discussed in Methods. 
The three-step Ramsey sequence is useful for mapping out the energy dispersion, $h f_Q$, over a wide range of $\varepsilon$, yielding the results shown with black dots in Figure~1c.  
This energy dispersion can be understood with the following three-level, hybrid qubit Hamiltonian:
\begin{equation}
H_{\rm eff}=
\begin{pmatrix}
\varepsilon/2 & \Delta_1 & \Delta_2 \\
\Delta_1  & -\varepsilon/2 & 0 \\
\Delta_2 & 0 & -\varepsilon/2+\Delta_R
\end{pmatrix},
\label{eq:H_eff}
\end{equation}
where the first basis state, $\ket{\mathcal{L}_0}$, has a singlet-like (2,1) charge configuration (two electrons in the left dot and one in the right), and the other two basis states, $\ket{\mathcal{R}_0}$ and $\ket{\mathcal{R}_1}$, have singlet-like and triplet-like (1,2) charge configurations.\cite{ShiPRL2012}
Here, $\Delta_1$ and $\Delta_2$ refer to the tunnel couplings between disparate charge states, and $\Delta_R$ is the energy splitting between the two (1,2) basis states, as indicated in Figure~1d.
The lowest two eigenstates of $H_\text{eff}$ correspond to the qubit levels $\ket{0}$ and $\ket{1}$, while the third state is an excited leakage level, $\ket{\tilde{\mathcal{L}}}$, as indicated in the inset of Figure~1c.
Fitting the experimental data to eq~1 yields the solid red line in the main panel, with (constant) fitting parameters $\Delta_1= 3.75$~GHz, $\Delta_2 =8.1$~GHz, and $\Delta_R = 12.25$~GHz.
The fit is quite good; 
however, eq~1 is a simple approximation, and deviations from this simple description can lead to significant, observable effects that are the focus of this paper.

\begin{figure}[t]
\includegraphics[clip,width=2.8 in]{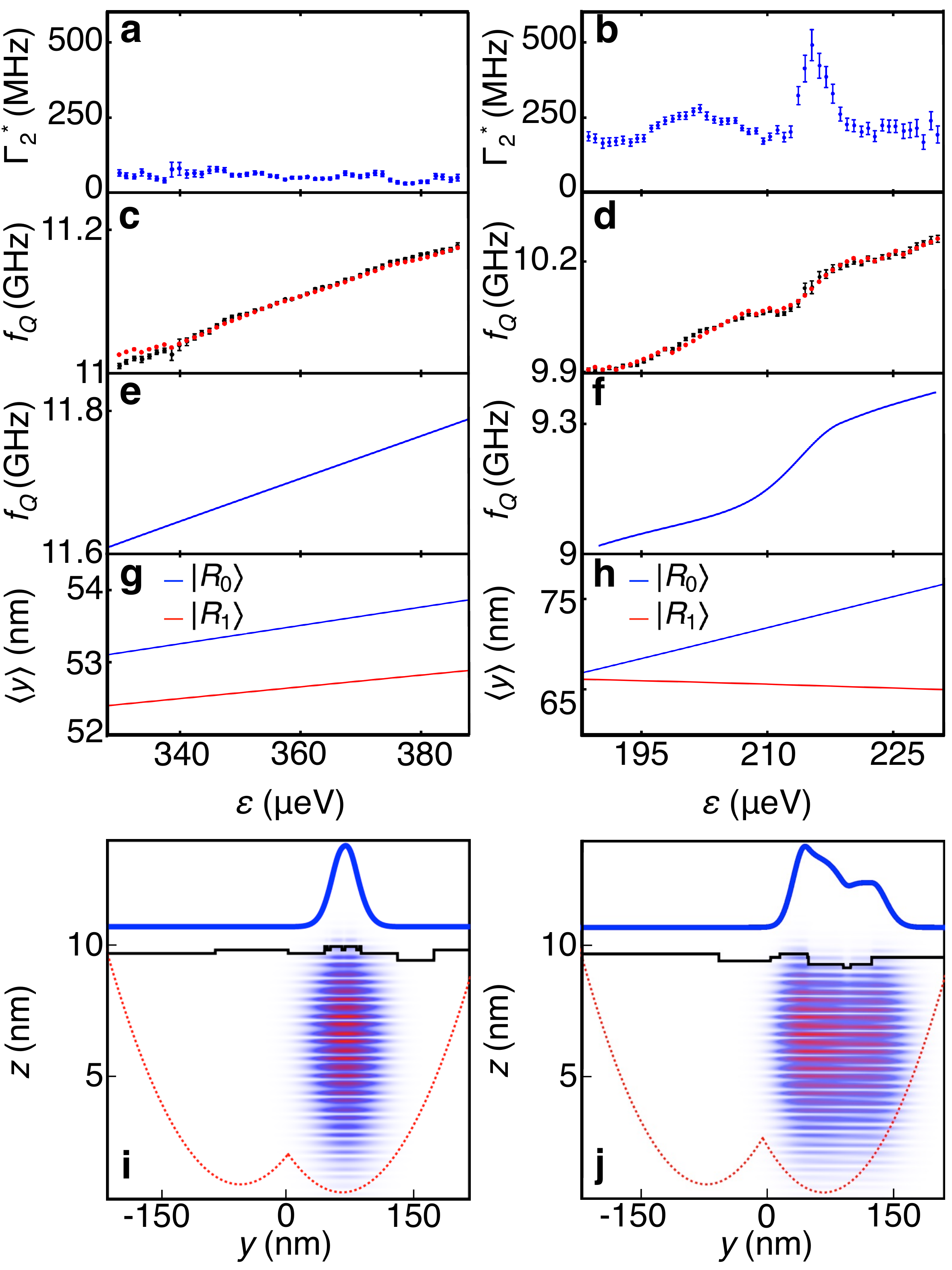}
\caption{
Experimental and theoretical analyses of normal (i.e., smooth) hybrid-qubit energy dispersions (left-hand column) vs.\ ``hot-spot" dispersions (right-hand column).
(a,b) Ramsey decay rates, $\Gamma_2^*$, obtained as in ref~\onlinecite{BrandurNPJ2017}. 
(c,d) Experimental measurements of $f_Q$ (black dots), plotted on the same horizontal axis as $\Gamma_2^*$.
The red dots are obtained by integrating the data in (a,b) with respect to $\varepsilon$, as described in Methods; the good agreement between black and red data, in both panels, shows that eq~2 is well satisfied, as expected when $\varepsilon$ fluctuations are the dominant decoherence mechanism.
The error bars in (a-d) are also discussed in Methods.
(e,f) $f_Q$, obtained from simulations, using the disorder profiles shown in (i,j).
Note that the magnitudes of $f_Q$ differ slightly, between experiments and simulations, but their relative variations are very similar.
(g,h) Centers of mass $\langle y\rangle$ for the ground (excited) qubit states, shown in blue (red).
(i,j) Lateral profiles of (1) the step disorder at the top quantum-well interface (black lines), (2) the double-dot confinement potential (red-dotted lines), and (3) the resulting charge density of the qubit ground state (heavy-blue lines).
2D plots of the ground-state charge density are also shown, with horizontal stripes corresponding to fast valley oscillations.
For the simulations, $z$=$0$ corresponds to the bottom quantum well interface.
Additional model parameters include
(e,g,i) $y_{R,L}$=$\pm 48.17$~nm, $F$=$1.2$~MV/m, $\hbar \omega$=$0.46$~meV;
(f,h,j) $y_{R,L}$=$\pm 67.85$~nm, $F$=$1.63$~MV/m, $\hbar \omega$=$0.38$~meV;
(i) $\varepsilon$=$360$~$\mu$eV; (j) $\varepsilon$=$210$~$\mu$eV.
(See Methods for an explanation of the various parameters.)}
\label{fig2}
\end{figure}

Figure~2a,b shows measurements of the dephasing rate $\Gamma_2^*$ for two tunings of the double dot that are different from each other and from that in Figure~1c.  ``Tuning'' here means a set of device gate voltages that determine $\Delta_1$, $\Delta_2$, and $\Delta_R$.  The tuning for Figure~2a shows little structure in $\Gamma_2^*$ as a function of $\varepsilon$, whereas that for Figure~2b reveals a large peak in this dephasing rate.
Figure~2c,d shows corresponding measurements of the qubit frequency $f_Q$ at these tunings, obtained using a conventional
Ramsey pulse sequence, as illustrated on the second line of Figure~1b. While $f_Q$ is a smooth function of $\varepsilon$ in Figure~2c, there is a step in $f_Q$ near $\varepsilon$=215~$\mu$eV in Figure~2d at the same location as the peak in $\Gamma_2^*$ in Figure~2b. 
Such a step clearly is inconsistent with eq~1 for detuning-independent Hamiltonian parameters, and its coincidence with the peak in $\Gamma_2^*$ is striking.

For solid-state qubits, charge noise is often the dominant decoherence mechanism.\cite{Vion886, TaylorPRL2013, MedfordPRL2013, KimNature2015, JiajiaPRB2015, ReedPRL2016, MartinsPRL2016, ShimPRB2016} In Ramsey measurements, the qubit phase evolves at a rate proportional to the qubit frequency $f_Q$, and the dephasing arising from charge noise obeys the relation\cite{PeterssonPRL2010, DialPRL2013}
\begin{equation}
\Gamma_2^*= \sqrt{2}\pi |\partial f_Q/\partial\varepsilon|\sigma_\varepsilon , 
\label{eq:Gamma2}
\end{equation}
where the standard deviation of the quasi-static charge noise, $\sigma_\varepsilon$, should be a constant for a given device, at a given temperature. 
Using this equation, we can integrate the $\Gamma_2^*$ data points in Figure~2a,b, as described in Methods, and compare the results to the measured $f_Q$ in Figure~2c,d, as shown by the red dots.
The correspondence between the integrated dephasing rate and $f_Q$ is remarkable, 
indicating that the step in $f_Q$ in Figure~2d indeed is converted by charge noise into a peak in the dephasing rate at that value of $\varepsilon$.

Figure~3 shows that atomic structure at the quantum well interface can also have a strong effect on Rabi oscillations.
Here, the data were obtained at a fixed driving frequency, corresponding to the qubit resonance condition near $\varepsilon$=225~$\mu$eV, and at a fourth overall tuning of the quantum device.
To determine the energy dispersion for a range of detunings about this value, we employ the Larmor pulse sequence shown in Figure~1b, yielding the results shown in Figure~3b.
In this case, the dispersion exhibits a maximum and a roughly 10~$\mu$eV plateau (a sweet spot) near $\varepsilon$=225~$\mu$eV, with sharp changes in the dispersion occurring on either side.
The long-lived Rabi oscillations near the dispersion plateau yield a decay rate of $\Gamma_{\mathrm{Rabi}}$=5.4~MHz, with much higher decay rates on either side of the plateau.
The dispersion-induced enhancement of the coherence time at this specific value of the detuning is also remarkable. 

For qubit gate operations, behavior like Figure~2a is clearly preferable to Figure~2b, and a sweet spot like Figure~3b would be optimal.
However, these different phenomena are not directly explained by eq~1 with conventional, constant parameters $\Delta_1$, $\Delta_2$, and $\Delta_R$.
We now argue that the unexpected behavior observed in the qubit energy dispersions can be explained by the presence of atomic-scale disorder at the upper quantum-well interface, which modifies the Hamiltonian model parameters due to interference between the Si conduction valleys.\cite{FriesenAPL2006,NestoklonPRB2006,KharcheAPL2007,FriesenPRB2010, CulcerPRB2010, Wu2012, GamblePRB2013,RohlingPRL2014,BorossPRB2016,GambleAPL2016}
To test this hypothesis theoretically, we consider a double-dot confinement potential for a single electron, as illustrated in Figure~1d.
Ignoring the excited state of the left dot, as appropriate when $\varepsilon$$\gg$0, the system can be described by the same three-level Hamiltonian as the quantum-dot hybrid qubit\cite{SchoenfieldNC2017} by replacing the three-electron basis with a one-electron basis comprised of a (1,0) charge configuration, $\ket{L_0}$, and two (0,1) charge configurations, $\ket{R_0}$ and $\ket{R_1}$.
The tunnel couplings $\Delta_1$ and $\Delta_2$ have the same meaning as before, while $\Delta_R$ corresponds to the low-energy splitting of the right dot, which could reflect a valley excitation, an orbital excitation, or a combination.\cite{FriesenPRB2010}
For a quantum-dot hybrid qubit, $\Delta_R$ also includes exchange and Coulomb terms; otherwise, the mapping between hybrid and charge qubits is exact.

\begin{figure}[t]
\includegraphics[clip,width=3.3 in]{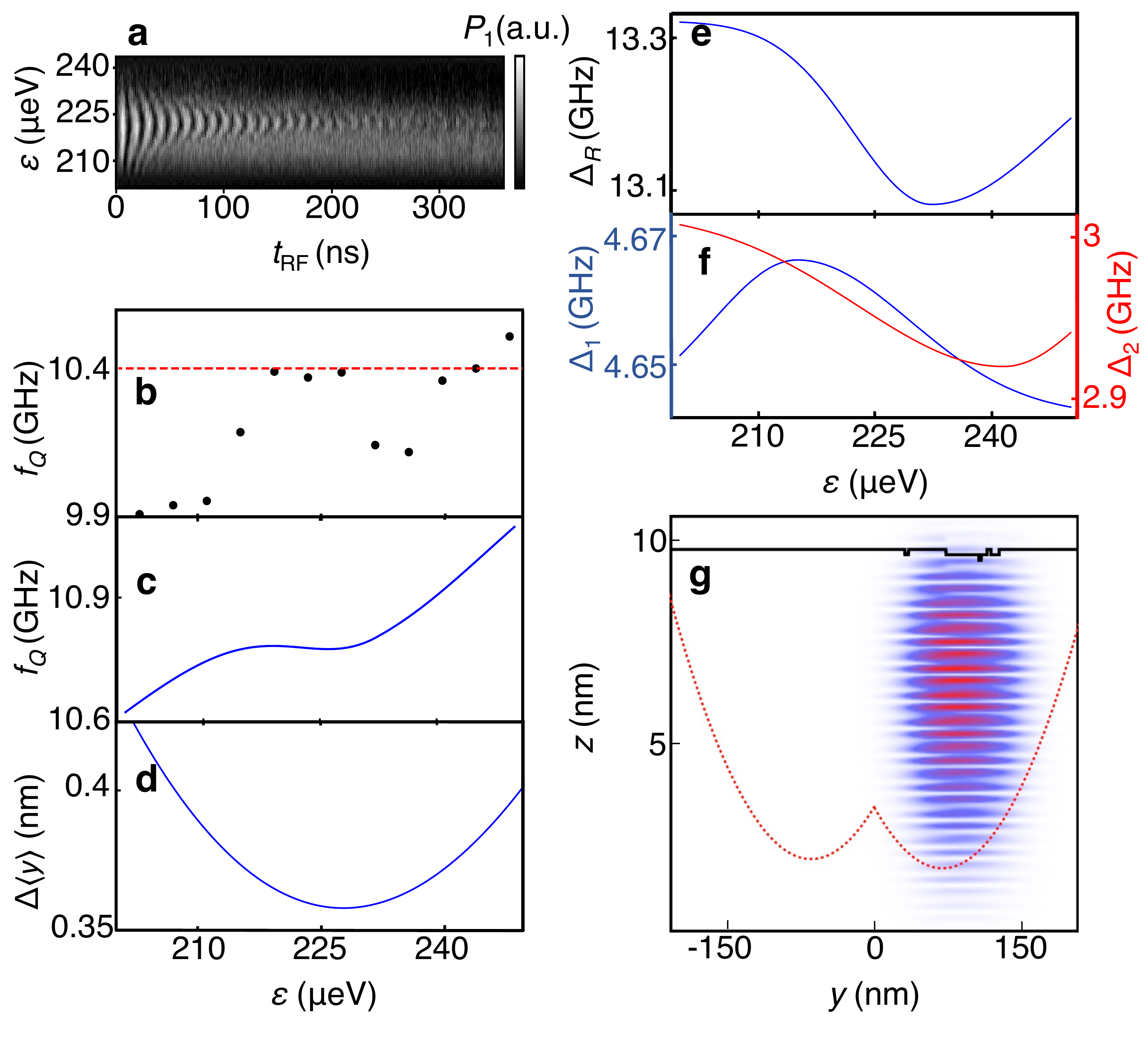}
\caption{Experimental and theoretical analysis of a ``sweet-spot" energy dispersion.
(a) Rabi fringes featuring an extended coherence region. 
Here, $P_1$ is the probability of being in state $\ket{1}$, and $t_\text{RF}$ is the duration of the microwave pulse.
(b) Experimental measurement of $f_Q$, based on a Larmor pulse sequence.
Here, the sweet spot occurs at the plateau near $\varepsilon$=225~$\mu$eV, and the red dashed line indicates the driving frequency used in (a).
(c) Qubit frequency $f_Q$, obtained from simulations, using the disorder profile shown in (g).
(d) The center of mass difference between the qubit states, defined as $\Delta \langle y\rangle$=$\langle y\rangle_1$$-$$\langle y\rangle_0$, plotted on the same horizontal axis as (b) and (c).
(e,f) The valley-splitting energy parameter, $\Delta_R$, and tunnel couplings, $\Delta_1$ and $\Delta_2$, obtained from simulations.
(See Methods.)
(g) Lateral profiles of (1) the interfacial disorder (black line), and (2) the double-dot confinement potential (red-dotted line).
A 2D plot of the ground-state charge density is also shown.
Simulation parameters are (c-g) $y_{R,L}$=$\pm 67.85$~nm, $F$=1.27~MV/m, $\hbar \omega$=0.42~meV;
(g) $\varepsilon$=225~$\mu$eV.
(See Methods for explanation.)}
\label{fig3}
\end{figure}

We simulate the effects of disorder in a single-electron double dot by constructing a minimal tight-binding model that captures the relevant valley physics.
As described in Methods, the Hamiltonian comprises terms describing the vertical quantum-well confinement (including atomic-scale disorder), the lateral double-dot confinement, the vertical electric field, and a lateral field representing the detuning.
The simulations assume Hamiltonian parameters consistent with the experiments.
In all cases, we consider a quantum well of width 9.85~nm and we focus on the ubiquitous atomic-step disorder arising from the underlying miscut of the substrate wafer, or from strain relaxation in the SiGe virtual substrate. 
Our results indicate that simple disorder profiles (e.g., single steps) are unable to explain the range of behaviors observed in the experiments. 
Moreover, we find that the effect of a given profile on the energy dispersion can be difficult to predict, \emph{a priori}. 
We have therefore performed a large number ($>$3,000) of simulations incorporating randomly generated step profiles, such as those shown in Figures~2i,j, and 3g. 
The disorder models we employ include steps ranging from 10 to 600 atoms in length, and we allow the position of the top interface, $z_t(y)$, to deviate from its average value by a standard deviation of 1 to 2 atoms. 
(See Methods for further details.) 
Other model parameters, including the positions of the left and right dots, the electric field, and the orbital excitation energy, are also chosen randomly, within a range of values consistent with our experiments.
After identifying promising configurations, we fine-tune the model parameters by hand to more closely match the experimental energy dispersions.
For simplicity, we do not include an overall miscut.

Disorder profiles that approximately replicate the normal, hot-spot, and sweet-spot behaviors are shown in Figures~2i,j, and 3g.
We allow for different disorder profiles in each of the simulations because the different tunings used in the experiments cause the dots to be exposed to different portions of the interface.\cite{ZhanAPL}
The resulting theoretical energy dispersions are shown in Figures~2e,f, and 3c, directly below their experimental counterparts.
Corresponding tight-binding wavefunctions are also shown in the figures, and we note that a significant amount of disorder is needed within the quantum dot to suppress the valley splittings to the levels observed in experiments; for comparison, disorder-free interfaces yield valley splittings $>$100~GHz.\cite{Boykin2004}

The hot spots and sweet spots reported here reflect the occasional occurrence of localized changes in typically smooth dispersions observed in both experiments and simulations.
By analyzing the simulation results, we can gain intuition into the origins of such exotic effects.
We have found that a comparison of the centers of mass (COM) $\langle y\rangle$ between the ground and excited valley states can be an effective indicator for unusual behavior.
For example, Figure~2g shows a typical COM response for a ``normal" (i.e., smooth) energy dispersion as a function of detuning.
Here, the COM of both eigenstates move smoothly and in tandem, displaying no distinctive features.  
This can be understood from Figure~2i, where we see that the wave function is centered at a location where it is not pressed against a step edge, resulting in no sudden changes as the detuning is varied.

On the other hand, the hot spot in Figure~2f has a very different COM response, as shown in Figure~2h.
Here, the two eigenstates are spatially well-separated (a valley-orbit coupling effect) and their positions are rapidly changing, which exposes them to distinct, local disorder potentials.
The valley composition of the eigenstates also varies rapidly, yielding sudden changes in the qubit frequency, as shown in Figure~2f.
An unexpected consequence of these effects is that for detunings around $\varepsilon$$\simeq$225~$\mu$eV the excited state $\ket{R_1}$ moves in opposition to the electric field, displaying a striking example of valley-orbit coupling.

To explain the sweet-spot behavior in Figure~3, we interpret the simulation results as follows.
Although the disorder profile of Figure~3g is jagged and rapidly varying, the COM of the qubit states shown in Figure~3d are closely spaced and move in tandem near the sweet spot.
Here we plot the relative COM, defined as $\Delta\langle y\rangle$=$\langle y\rangle_1$$-$$\langle y\rangle_0$.
Away from the sweet spot, the eigenstates move more independently.
Moreover, the valley splitting parameter $\Delta_R$, which dominates the qubit frequency in the far-detuned regime $\varepsilon$$\gg$0, also exhibits a minimum at the sweet spot (see Methods for details on extracting $\Delta_R$); when combined with the slowly increasing ``background" qubit frequency (e.g., Figure~2c), we obtain the relatively flat dispersion shown in Figure~3c.

In this letter, we have reported hot spots and sweet spots that are not anticipated by the usual models describing quantum-dot qubits.
We attributed these features to atomic-scale disorder at the quantum-well interface, and showed that they can directly affect the dephasing of a quantum-dot hybrid qubit.
To clarify the physics, we performed tight-binding simulations of a double dot, taking into account both conduction-band valleys and the valley-orbit coupling caused by step disorder.
By introducing random disorder profiles, we were able to generate dispersion features consistent with the experiments.
In both theory and experiment, in most cases, we observed no distinct features in the dispersion.  
However, special disorder profiles were found to induce hot spots (or sweet spots), where the qubit is particularly susceptible to (or protected from) electrical fluctuations of the detuning parameter.
Since atomic-scale disorder is ubiquitous in Si heterostructures, these results suggest that it could be possible to enhance quantum coherence in future Si qubit experiments by electrostatically tuning the dots so they are exposed to desirable disorder profiles.

\textbf{Methods.} 
\emph{Measuring the energy dispersion in Figure~1c.}
Following ref~\onlinecite{BrandurNPJ2017}, we first initialize the qubit into its ground state, $\ket{0}$, at the north pole of the Bloch sphere.  
We then apply the 3-step pulse sequence, illustrated in Figure~1b, which allows us to measure the qubit frequency over a wide range of detunings.
A microwave voltage pulse corresponding to an $X_{\pi/2}$ rotation is applied to the gate labeled R in Figure~1a in order to rotate the qubit onto the equator of the Bloch sphere.  
The dc bias voltage on gate~R is then adiabatically adjusted to give the desired detuning $\varepsilon$.
Free induction ensues for a time period, $t_\text{free}$, after which the detuning is adiabatically returned to its initial value, and a second $X_{\pi/2}$ rotation is performed.  
The qubit is then measured to determine the probability $P_1$ of being in the excited state, $\ket{1}$, at the south pole of the Bloch sphere, and the experiment is repeated as a function of $t_\text{free}$, to obtain Ramsey fringes.  By Fourier transforming these data, we determine the qubit frequency $f_Q$ corresponding to $\varepsilon$.  The experiment is then repeated, keeping all parameters fixed except $\varepsilon$, to obtain a map of the energy dispersion.

\emph{Measuring the energy dispersions in Figure~2.}
Here we follow the same procedure as Figure~1, replacing the 3-step pulse with a conventional Ramsey pulse sequence, as illustrated in the second line of Figure~1b.
To determine the Ramsey decay rates, $\Gamma_2^*$, shown in Figure~2a,b, we fit the Ramsey fringes to an exponentially decaying sinusoid function.\cite{BrandurNPJ2017}
The error bars in Figure~2a-d were obtained from the covariance matrix determined during this procedure.

\emph{Measuring the energy dispersions in Figure~3.}
In Figure~3a, we use the Rabi pulse sequence illustrated in Figure~1b, applied to gate~L.
In this case, the frequency of the oscillations depends on the microwave power, rather than the qubit energy splitting.
The Rabi decay rate reported in the main text is obtained by fitting the Rabi oscillations to an exponentially decaying sinusoid at the $\varepsilon$ value corresponding to the slowest Rabi oscillations.

In Figure~3b, we apply the Larmor pulse sequence illustrated in Figure~1b and described in ref~\onlinecite{ShiNature2014} to gate~L.
In this case, after initialization, the qubit is abruptly pulsed to a desired value of $\varepsilon$, putting it in a superposition of qubit eigenstates. 
Free induction ensues for a time period, $t_\text{free}$, after which the detuning is abruptly pulsed back to its initial value where the qubit is measured. 
Repeating the experiment as a function of $t_\text{free}$ yields Larmor fringes, which are Fourier transformed, analogous to the Ramsey experiment, to obtain the energy dispersion.

\emph{Numerical integration of eq 2.}
The red dots in Figure~2c,d were obtained by numerically integrating the data in Figure~2a,b.
If we use the indices $i$ ($j$) to label the $i$th ($j$th) data points for $\Gamma_2^*$ and $f_Q$, and note that the distance between detuning steps is a constant, $\Delta \varepsilon$, then the numerical integral can be expressed as
\begin{equation}
\tilde{f}_{Q,i}=\tilde{f}_{Q,0}+(\sqrt{2}\pi\sigma_\varepsilon\Delta\varepsilon)\sum_{j=0}^i\Gamma_{2,j}^* \text{sign} [f_{Q,j}-f_{Q,j-1}] ,
\end{equation}
where $i>0$, and $\tilde{f}_Q$ is the integrated estimate for $f_Q$. 
Note that the function $\text{sign} [f_{Q,j}$-$f_{Q,j-1}]$ accounts for the absolute value sign in eq~2, and is evaluated using experimental data.
In Figure~2c,d, we use the same value of $\sigma_\varepsilon=4.39$~$\mu$eV, which is also consistent with ref~\onlinecite{BrandurNPJ2017}.

\emph{Tight-binding model.}  
For a strained Si quantum well, the two low-lying conduction band valleys are centered at positions ${\bf k}_0$=$\pm 0.82(2\pi/a)\hat{\bf z}$ in the Brillouin zone, where $a$=0.543~nm is the length of the (unstrained) Si cubic unit cell, and $\hat{\bf z}$ is the growth direction, which we assume here to be oriented along (001), 
for simplicity.
The minimal tight-binding model captures these valley positions as well as their longitudinal and transverse effective masses ($m_l$=$0.916\, m_0$ and $m_t$=$0.191\, m_0$, respectively) by introducing nearest- and next-nearest-neighbor hopping parameters in the $z$~direction~\cite{Boykin2004,Boykin2004PRB} ($u_z$=0.68~eV and $v_z$=0.61~eV, respectively), and a separate nearest-neighbor hopping parameter in the $x$-$y$ plane \cite{Shiau2007,SaraivaPRB2010,GamblePRB2013} ($u_y$=$-10.91$~eV).
The double-dot confinement potential in the $x$-$y$ plane is obviously three-dimensional (3D).
However, an interfacial step is a 2D feature that generates valley-orbit coupling in the $x$-$y$.\cite{FriesenPRB2010}
If we define the step direction as $\hat{\bf x}$, and further orient the double-dot axis along $\hat{\bf y}$, then the essential physics of our problem is all contained within the $y$-$z$ plane, and inclusion of the third dimension ($\hat{\bf x}$) only provides quantitative corrections, but no new physics.
Our minimal model can therefore be reduced to the $y$-$z$ plane.

The hopping parameters, described above, account for the kinetic energy, $H_K$, of an electron in a strained-Si quantum well.
The electronic potential energy is described via on-site (i.e., diagonal) terms, involving several contributions.
(1) We include a uniform on-site energy of 23.23~eV, which ensures a ground-state energy of zero for an infinite-size system with no other confining potentials or fields.
(2) We introduce a quantum well with a barrier of height $V_\text{QW}$=0.15~eV, as appropriate when Si is sandwiched between strain-relaxed Si$_{0.7}$Ge$_{0.3}$.\cite{Schaeffler}
If we define the position of the bottom interface of the well as $z_b=0$, and assume the top well interface $z_t(y)$ is a function of position (i.e., the steps), then the barrier potential can be written as
\begin{equation}
H_\text{QW}=V_\text{QW}\left[\theta(z_b-z)+\theta(z-z_t(y))\right] ,
\label{eq:H_SiGe}
\end{equation}
where $\theta(z)$ is the Heaviside step function.
(3) We include a vertical electric field $F$, as consistent with experiments, which pulls the electron wavefunction up against the top interface:
\begin{equation}
H_F=-eFz .
\label{eq:Fz}
\end{equation}
Ideally, this field should be large enough that the electron feels no confinement effects from the bottom of the quantum well.
(We note that electric fields in the range of $F$=1-2~MV/m, which were reported in Figures~2 and 3, satisfy this criterion. However, we have also observed good results at higher fields, of order 6~MV/m.)
(4) We model the two dots, centered at positions $y_L$ and $y_R$, with a biquadratic potential:
\begin{equation}
H_\text{DD}=\min\left[\frac{1}{2}m_t\omega^2(y-y_L)^2,\frac{1}{2}m_t\omega^2(y-y_R)^2\right] ,
\label{eq:H_DD}
\end{equation}
where $\omega$ represents the orbital excitation frequency of the individual dots. 
For simplicity, we assume both dots have the same $\omega$.
(5) We include the effects of a detuning parameter $\varepsilon$ via an in-plane electric field:
\begin{equation}
H_{\varepsilon}=-\frac{\varepsilon}{2(y_R-y_L)} y .
\label{eq:H_det}
\end{equation}
The full Hamiltonian of the system is then written as
\begin{equation}
H=H_{K}+H_\text{QW}+H_{F}+H_\text{DD}+H_{\varepsilon}.
\label{eq:H_total}
\end{equation}

\emph{Fitting $\Delta_1$, $\Delta_2$, and $\Delta_R$ in simulations.}  
In Figure~1, the Hamiltonian parameters $\Delta_1$, $\Delta_2$, and $\Delta_R$ were determined by fitting the experimental data in Figure~1c to eq~1, assuming the fitting parameters to be independent of $\varepsilon$.
This is a good approximation for ``normal" dispersion relations, which are smooth, with no distinct features.
The approximation is not good for sweet spots or hot spots.
Below, we describe our method for extracting $\Delta_1$, $\Delta_2$, and $\Delta_R$ as a function of $\varepsilon$ from the simulation results, as shown in Figure~3e,f.

We consider only the far-detuned regime, $\varepsilon$$\gg$0, where the two low-energy eigenstates have charge configuration (0,1).
The key is to determine the valley splitting, $\Delta_R$, independently of $\Delta_1$ and $\Delta_2$, by making the following approximation:
we replace the double-dot confinement potential, eq~6, with the right-localized single-dot potential,
\begin{equation}
H_\text{SD}=\frac{1}{2}m_t\omega^2(y-y_R)^2 ,
\end{equation}
and repeat the tight-binding simulation, assuming the same interfacial disorder potential.
Repeating this procedure as a function of $\varepsilon$ gives $\Delta_R(\varepsilon)$.
Ignoring the left dot in this way is a good approximation because the tails of the wave function do not play a significant role in determining the valley splitting.
On the other hand, the tails play an important role in determining the tunnel couplings $\Delta_1(\varepsilon)$ and $\Delta_2(\varepsilon)$.
We compute these quantities by solving the roots of the characteristic polynomial of eq~1 at each $\varepsilon$, using the previously computed function $\Delta_R(\varepsilon)$.

\vspace{0.2cm}\noindent\textbf{ACKNOWLEDGMENTS}

\noindent Work in the U.S. was supported in part by ARO (W911NF-17-1-0274, W911NF-12-0607, W911NF-08-1-0482), NSF (DMR-1206915, PHY-1104660, DGE-1256259), and the Vannevar Bush Faculty Fellowship program sponsored by the Basic Research Office of the Assistant Secretary of Defense for Research and Engineering and funded by the Office of Naval Research through Grant No.\ N00014-15-1-0029. 
Work in Spain was supported in part by MINEICO and FEDER funds through Grant Nos.\ FIS2012-33521, FIS2015-64654-P, BES-2013-065888, EEBB-I-17-12054, and by CSIC through Grant No. 201660I031.
Work in South Korea was supported by the  Korea Institute of Science and Technology Institutional Program (Project No.\ 2E26681).
Development and maintenance of the growth facilities used for fabricating samples is supported by DOE (DE-FG02-03ER46028). We acknowledge the use of facilities supported by NSF through the UW-Madison MRSEC (DMR-1121288). The views and conclusions contained in this document are those of the authors and should not be interpreted as representing the official policies, either expressed or implied, of the Army Research Office (ARO), or the U.S. Government. The U.S. Government is authorized to reproduce and distribute reprints for Government purposes notwithstanding any copyright notation herein. 

\bibliography{disorder}

\end{document}